\documentclass[prl,amsmath,aps,twocolumn,superscriptaddress]{revtex4}
\usepackage{graphicx}
\pdfoutput=1
\usepackage{color}
\begin{document}
\newcommand{\red}{\color{red}}
\newcommand{\scalar}[2]{\left \langle#1\ #2\right \rangle}
\newcommand{\me}{\mathrm{e}}
\newcommand{\mi}{\mathrm{i}}
\newcommand{\dif}{\mathrm{d}}
\newcommand{\period}{\text{per}}
\newcommand{\free}{\text{fr}}
\newcommand{\rme}{\mathrm{e}}
\newcommand{\rmd}{\mathrm{d}}
\newcommand{\nn}{\nonumber}
\renewcommand{\epsilon}{\varepsilon}
\newcommand{\mq}[2]{\uwave{#1}\marginpar{#2}}
\newcommand{\fig}[2]{\includegraphics[width=#1\columnwidth]{./#2}}
\newcommand{\Fig}[1]{\includegraphics[width=\columnwidth]{./#1}}
\newcommand{\FFig}[1]{\includegraphics[width=0.87\columnwidth,angle=270]{./#1}}
\newlength{\bilderlength}
\newcommand{\bilderscale}{0.35}
\newcommand{\storebilderscale}{\bilderscale}
\newcommand{\bilderskip}{\hspace*{0.8ex}}
\newcommand{\textdiagram}[1]{%
\renewcommand{\bilderscale}{0.25}%
\diagram{#1}\renewcommand{\bilderscale}{\storebilderscale}}
\newcommand{\diagram}[1]{%
\settowidth{\bilderlength}{\bilderskip%
\includegraphics[scale=\bilderscale]{#1}\bilderskip}%
\parbox{\bilderlength}{\bilderskip%
\includegraphics[scale=\bilderscale]{#1}\bilderskip}}
\newcommand{\Diagram}[1]{%
\settowidth{\bilderlength}{%
\includegraphics[scale=\bilderscale]{#1}}%
\parbox{\bilderlength}{%
\includegraphics[scale=\bilderscale]{#1}}}

\graphicspath{{figures/},{}}

\bibliographystyle{../macros/KAY.bst}
\definecolor{c1}{rgb}{1, 0, 0}
\definecolor{c2}{rgb}{0, 1, 0}
\definecolor{c3}{rgb}{0, 0, 1}
\definecolor{c4}{rgb}{1, 0, 1}
\definecolor{c5}{rgb}{0, 1, 1}

\title{Critical interfaces in the  random-bond Potts model}
\author{Jesper L.~Jacobsen}
  \affiliation{CNRS-Laboratoire de Physique Th\'eorique de l'Ecole Normale
  Sup\'erieure, 24 rue Lhomond, 75231 Paris, France.}
\author{Pierre Le Doussal}
\affiliation{CNRS-Laboratoire de Physique Th\'eorique de l'Ecole Normale
  Sup\'erieure, 24 rue Lhomond, 75231 Paris, France.}
  \author{Marco Picco}
  \affiliation{CNRS-LPTHE, Universit\'es Paris 6 et Paris 7, 4 Place Jussieu, 75005 Paris, France}
\author{Raoul Santachiara}
\affiliation{CNRS-Laboratoire de Physique Th\'eorique de l'Ecole Normale
  Sup\'erieure, 24 rue Lhomond, 75231 Paris, France.}
  \author{Kay J\"org Wiese}
  \affiliation{CNRS-Laboratoire de Physique Th\'eorique de l'Ecole Normale
  Sup\'erieure, 24 rue Lhomond, 75231 Paris, France.}

\begin{abstract}
  We study geometrical properties of interfaces in the random-temperature $q$-states Potts model as an example of a conformal
  field theory weakly perturbed by quenched disorder.  Using conformal
  perturbation theory in $q-2$ we compute the fractal dimension of
  Fortuin Kasteleyn domain walls. We also compute it numerically both
  via the Wolff cluster algorithm for $q=3$ and via transfer-matrix
  evaluations. We also obtain numerical results for the fractal
  dimension of spin clusters interfaces for $q=3$. These are found
  numerically consistent with the duality $\kappa^{\mathrm{spin}}
  \kappa ^{\mathrm{FK}}= 16$ as expressed in putative SLE parameters.
\end{abstract}

\maketitle


The discovery of the Schramm L\"owner Evolution (SLE) has strongly revived interest in geometrical
properties of interfaces in 2-dimensional statistical physics. SLE provides a rigorous classification, with a
single parameter $\kappa$, of probability measures on non-crossing random fractal curves, which satisfy both
conformal invariance  and the domain Markov property \cite{schramm00}. Interfaces and similar
geometric objects defined in pure 2-dimensional critical models are conjectured, and in some cases proven, to
satisfy both requirements in the continuum limit. SLE hence describes such diverse systems as percolation
$\kappa=6$, self-avoiding walks $\kappa=4/3$, loop-erased random walks $\kappa=2$ and level lines of height
models $\kappa=4$ \cite{sleBernardBauerRev06
}. It applies to the Ising and 3-states Potts
interfaces, both for spin clusters ($\kappa=3$ and $\kappa=10/3$, respectively) and the dual Fortuin
Kasteleyn (FK) clusters ($\kappa=16/3$ and $\kappa=24/5$), with a duality $\kappa \leftrightarrow
\kappa'=16/\kappa$. While these models have been described, prior to SLE, using conformal field theory (CFT),
SLE bridges the gap between the algebraic approach of CFT and the geometry of interfaces. The present connections between SLE and CFT \cite{sleBernardBauerRev06
} focus on boundary-condition changing operators, which generate the curves. 
They give 
$d_{\mathrm{f}}=1+\kappa/8$ for the fractal dimension of
the interface, i.e.\ the hull of the SLE trace. Extensions beyond non-minimal CFT \cite{nonminimal}  are rare. 

A tantalizing question is whether CFT methods and SLE help to understand a broader class of scale-invariant 2d
complex systems, such as systems with quenched disorder or far from equilibrium. Numerical
studies indicate that zero-vorticity lines in 2D-turbulence \cite{turbulence} and domain walls in
spin-glasses \cite{spinglass} may be described by SLE. 
These examples are ``far'' from any pure CFT, thus the situation may be more
favorable for models which are ``weak perturbations'' of a known CFT. This is  e.g.\ the case for the $q$-states
Potts model, perturbed by quenched random bond (i.e., temperature) disorder, known to exhibit a stable weak-disorder fixed point for $q>2$, perturbatively accessible in a $q-2$ expansion. This has been studied
 using conformal perturbation theory \cite{Ludwig1987,DotsenkoPiccoPujol1995} and transfer-matrix
methods \cite{randompotts}. However, geometric properties of interfaces which are crucial for future comparison to
SLE
 were to our knowledge not
investigated. 

The aim of this Letter is to present results for the fractal
dimension of domain walls in the random-temperature Potts
model. These are obtained by analytical calculation using
conformal perturbation theory inspired by
\cite{Ludwig1987,DotsenkoPiccoPujol1995}, and
from two types of large-scale numerics: Monte Carlo
simulations using the efficient Wolf-algorithm \cite{Wolff1989}, which
allow to keep track of both spin and FK clusters in the same
simulation, and transfer-matrix calculations, whose advantage is to
make close contact with CFT. The results of all
three methods agree nicely.

Let us recall the definition of the model: In terms of the spin
variables $\sigma_i=\{1,\cdots,q\}$ at lattice site
$i$, the partition function of the $q$-states Potts model is
\begin{equation*}\label{def_potts_spin}
{\cal Z} = \sum_{\{\sigma_{i}\}}  \rme^{\beta  \sum_{\left<i j\right>} J_{ij} \delta_{\sigma_{i}\sigma_{j}} }
\sim \sum_{\{\sigma_{i}\}}  \prod_{\left<ij\right>} \left[1-{\sf p}_{ij}+ {\sf p}_{ij} \delta_{\sigma_{i}\sigma_{j}} \right]\ ,
\end{equation*}
where the sum runs over nearest-neighbor bonds 
$\left< i,j \right>$. The last expression is the spin-cluster
expansion, noting $1-{\sf p}_{ij}= \rme^{-\beta J_{ij}}$. By expanding in
${\sf p}_{ij}$, it can be rewritten in terms of  the
Fortuin-Kasteleyn (FK) clusters, composed by placing a bond between
neighboring sites with probability ${\sf p}_{ij}$. The pure ferromagnetic
model has $J_{ij}=J>0$, ${\sf p}_{ij}={\sf p}$, while in the disordered
one the $J_{ij}$ are chosen as i.i.d.\ random variables.
The partition function in the FK
representation is (up to a prefactor)
\begin{equation}\label{def_potts_fk}
{\cal Z} \sim \sum_{\cal G} {\sf p}^{|{\cal G}|} (1-{\sf p})^{|\overline {\cal G}|} q^{|| {\cal G}||}\ , \quad
\end{equation}
for the pure model, with a straightforward generalization to the
random case. Here $\mathcal{G}$ runs over all clusters (i.e.\ domains
connected by the above placed bonds), $\left|{\cal G} \right|$ is the
number of bonds, $\left|\overline{\cal G} \right|$ the not placed
bonds, and $||{\cal G}||$ the number of connected components. The partition sum (\ref{def_potts_fk}) allows to define the
Potts model with non-integer $q\geq0$. For the pure model it has a
continuous  phase transition for $0 \leq q \leq 4$, which becomes first order for  $q>4$.

Our analytical calculation focuses on weak disorder, where the
$J_{ij}= {\overline{J}}+\delta J_{ij}$ are Gaussian random variables
of variance $\beta^2 \overline{\delta J_{ij}^2}=g_{0}$ and
$\sqrt{g_{0}}\ll \beta \overline{J}$. Near the critical temperature of
the pure model, the continuum limit of the random Potts model can be
written \cite{Ludwig1987,DotsenkoPiccoPujol1995} as 
${\cal H} = {\cal H}_{\mathrm{pure}} + \int_{\vec z} \epsilon (\vec z) \delta J (\vec z)$
where $\int_{\vec z} \equiv  \int \rmd ^{2} \vec z$ and 
$\beta {\cal H}_{\mathrm{pure}}$ is the action of the pure $q$-state
Potts model, which at criticality can be identified with its conformal
field theory, or the $O(N=\sqrt{q})$ model
\cite{Nienhuis1984}. We use the Coulomb-gas representation of the
latter.  The coupling constant $p$ is related to $q$ via
$\sqrt{q}=2\cos (\pi/(2p))$, 
so that $p=2$ is Ising and $p=3$ is  3-state Potts. The second
term in ${\cal H}$  is the deviation from the pure
critical point induced by the disorder, where $\varepsilon(\vec z)$ is the
energy density operator of the pure model.
To  average over  disorder, 
the $n$-times replicated action is taken:
\begin{align}\label{}
&\ln \overline{e^{- \beta \sum _{a=1}^{n} {\cal H}^{a}}} = - \beta \sum _{a=1}^{n} {\cal H}_{\mathrm{pure}}^{a}
 + g_{0} \int_{\vec z}  \sum _{a,b=1}^{n}\epsilon^{a}(\vec z)\epsilon^{b}(\vec z) \nonumber \ , 
\end{align}
where everywhere below we use the shorthand notation $\Phi(z)$ to denote $\Phi(\vec z)$,
where $\vec z=(z,\bar z)$. The diagonal term
$\epsilon^{a} (z)^{2}$ is perturbatively less relevant
\cite{Ludwig1987} than $\epsilon^{a} (z)\epsilon^{b} (z)$, whose
dimension is $4 \Delta_{\epsilon }$, i.e., four times the dimension of the
holomorphic part of the energy primary field $\epsilon (z)\equiv
\Phi_{12} (z)$, with $\Delta_{\epsilon } =\frac{p+1}{2 (2p-1)}$. For
the Ising model, $p=q=2$, and disorder is  marginally irrelevant,
whereas for the 3-states Potts model $p=q=3$ it is relevant. Since the
Coulomb gas is defined for all $p$, we can perturbatively expand
around the Ising model \cite{Ludwig1987}. This expansion is
conceptually the same as for the $\phi^{4}$ model, except that Feynman
diagrams are evaluated using the unperturbed CFT (with averages denoted
$\langle ... \rangle_0$).  
We keep the perturbed system
on its critical manifold, s.t.\ only the renormalization of the disorder $g_{0}$ is
left to consider, with a correction to second order $O(g_0^2)$:
\begin{align}\label{2.2}
&\textstyle \sum_{a\neq b}\epsilon^{a} (z) \epsilon^{b} (z)  \sum_{c\neq
d}\epsilon^{c} (z') \epsilon^{d} (z') \nonumber \\
&\textstyle= 4 (n-2) \sum_{b\neq d} \epsilon^{b} (z) \epsilon^{d}
(z) \left< \epsilon (z)\epsilon (z') \right>_{0} +\dotsb\ .
\end{align}
Using the unperturbed average $\left< \epsilon
  (z)\epsilon (z') \right>_{0} = \frac{1}{|z-z'|^{4
    \Delta_{\epsilon}}}$ one obtains the renormalized disorder
    $g L^{4\Delta_{\epsilon}-2}=g_{0}+ 4 \pi  (n-2) g_{0}^{2}  \frac{L^{2-4\Delta_{\epsilon}}}{2-4\Delta_{\epsilon}}$, $L$ being the infrared cutoff, and 
%
%
the $\beta$-function (for $q>2$) \cite{Ludwig1987}:
\begin{equation}\label{lf49}
L\partial_{L} g = (2-4\Delta_{\epsilon}) g  + 4 \pi (n-2) g^{2} + \dotsb
\end{equation}
At $n=0$, $\beta(g)$ has an infrared fixed point at $g^{*} =
\frac{1-2\Delta_{\epsilon}}{4\pi}$ which determines the low-energy
behavior of the random model.  Conformal symmetry is expected to be
restored at 
$g^{*}$. To date, this method has
been employed to calculate the scaling dimension of the energy density
$\epsilon$ and of the spin $\sigma$, to two- and three-loop order in
\cite{Ludwig1987,DotsenkoPiccoPujol1995} and
\cite{DotsenkoPiccoPujol1995} respectively. The multi-scaling
properties of spin-spin correlation function has been determined in
\cite{Ludwig1990_DDP_97_Lewis98}.

Here we focus on geometrical properties, hence on the operator $\Phi_{10}(z)$ 
which measures \cite{RushkinBettelheimGruzbergWiegmann2007
} the passage of one critical curve at point $\vec z$. Indeed, for the pure model, the correlation function 
$\left<\Phi_{10}(z)\Phi_{10}(0)\right>_0=|z|^{-4\Delta_{10}}$ gives
the probability that two points lie at the perimeter of the same FK
cluster, from which one obtains the fractal dimension of FK domain walls $d_{\mathrm{f}}^{\mathrm{FK,pure}} = 2 - 2 \Delta_{10}$,
i.e. $d_{\mathrm{f}}^{\mathrm{FK}} =8/5$ for $q=3$. Here we  compute the corresponding
probability for the disordered system. A crucial question is 
whether $\Phi_{10} (z)$ is still the ``curve-detecting'' operator in
the disordered system. This is true at the ``critical
dimension'' $p=q=2$.  Increasing $p$ deforms the operator
adiabatically. Since the latter  is a physical
observable, it is an eigenoperator of the RG. We must check if there is an operator at $p=2$  which (i) has  the same dimension as $\Phi_{1,0}$, and (ii) appears in the  sub-algebra generated  by $\Phi_{1,0}$ and $\Phi_{1,2}$. If such an operator exists, it 
mixes with $\Phi_{1,0}$, and the curve-detecting operator
will be one of the eigenoperators of the RG flow involving $\Phi_{10}$.  We checked the absence of such an operator
: thus, at least for small $p-2$, $\Phi_{10}$ is
the curve-detecting operator. 

We now sketch the calculation of the scaling dimension of $\Phi_{10}$, for
details see \cite{long1}. There is no contribution to
order $g_0$, since contracting the disorder operator $\sum_{b\neq
  c}\epsilon^{b}\epsilon^{c}$ with $\Phi_{10}$ in, say, replica $a$,
leaves one $\epsilon$ in replica $b\neq a$, thus is not
proportional to $\Phi_{10}^{a}$. At 
second order, contracting two disorder vertices with
$\Phi_{10}^{a}(z_{1})$ gives
\begin{equation*}
\Phi_{10}^{a}(z_{1})   \frac{g_{0}^{2}}{2!}\bigg[ \sum_{b\neq c} \int_{z_{2}}
\epsilon^{b}(z_{2})\epsilon^{c}(z_{2})\bigg] \bigg[\sum_{d\neq e} \int_{z_{3}} \epsilon^{d}(z_{3})
\epsilon^{e}(z_{3})\bigg]
\end{equation*}
and projecting onto $\Phi_{10}^{a}(z_{1})$. Contracting using $\left< \epsilon^{c}
  (z_{2})\epsilon^{e} (z_{3}) \right>_0 = \delta^{ce}
|z_{2}-z_{3}|^{-4\Delta_{12}}$ to eliminate replicas not equal to $a$
we obtain  $\Phi_{10}^{a}(z_{1}) \epsilon^{b=a}(z_{2})\epsilon^{d=a}(z_{3})$,
which, projected onto $\Phi_{10}^{a}(z_{1})$ yields
%
\begin{equation}\label{lf53}
{{}_{z_{1}}\!\diagram{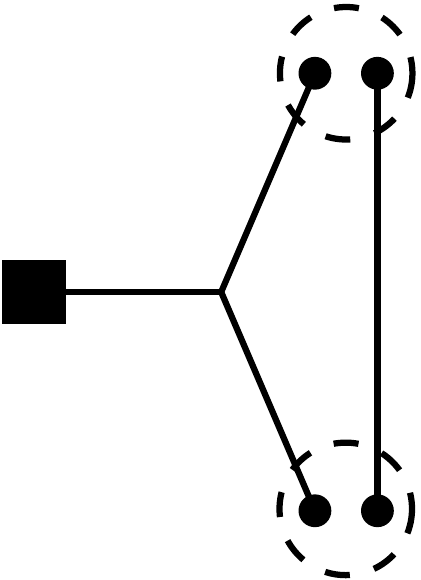}_{\!z_{2}}^{\!z_{3}}} =
\begin{array}{l}
\displaystyle
\frac{g_{0}^{2}}{2!} 4
(n-1) \int_{z_{2},z_{3}} \left<\epsilon\nopagebreak
(z_{2})\epsilon (z_{3})  \right>_{0}  \\
\displaystyle
\times \left(\Phi_{10} (z_{1})\epsilon
(z_{2})\epsilon (z_{3}) \big| \Phi_{10} (z_{1}) \right)\ ,
\end{array}
\end{equation}
where the OPE coefficient
$
 \big(\Phi_{10} (z_{1})\epsilon
(z_{2})\epsilon (z_{3}) \big| \Phi_{10} (z_{1}) \big)\!:=\lim_{R\to\infty}  \frac{\left< \Phi_{10} (z_{1})\epsilon
(z_{2})\epsilon (z_{3})  \Phi_{10} (R)  \right>_0}{\left< \Phi_{10} (z_{1})  \Phi_{10} (R)  \right>_0}
$.
%
This and the integral (\ref{lf53}) are computed using Coulomb gas
techniques \cite{DotsenkoCFT}. One 2d integration,
over one angle and one scale, is easy, and gives a pole in $1/
(p-2)$.  One 2d integral over say $z_{2}$ is left,
but we also need a screening charge $V_{+}$ to get the 4-point
function in (\ref{lf53}). 
We evaluated this
integral in the marginal dimension, i.e.\ for $p=2$
(Ising) by analytical techniques 
\cite{DotsenkoPiccoPujol1995}, and
numerically  \cite{long1}. The
result is 
\begin{align}\label{}
&\int_{z_{2},z_{3}}\left (\Phi_{10} (z_{1})\,\epsilon (z_{2})\,\epsilon
(z_{3}) \Big|\Phi_{10}(z_1) \right)  \left<\epsilon (z_{2}) \epsilon (z_{3})\right> \nn \\
&=-7.0710
{ L^{4-8\Delta_{12}}}{ (1-2\Delta_{12})^{-1}}\ ,
\end{align}
\nopagebreak
Inserting the fixed-point value $g^{*}$ from above gives
$
\dim_{L} (\Phi_{10}) = -2\Delta_{10} +
\frac{(1-2\Delta_{\epsilon})^{2}}{2\pi^{2}} \times 7.071
\stackrel{p=3}{=} -\frac{2}{5} + 0.01433
$.
%
%
This leads to the fractal dimension of FK domain walls:
\begin{equation}\label{16}
d_{\mathrm{f}}^{\mathrm{FK}} =
2+ \dim_{L} (\Phi_{10}) = 1.61433\ .
\end{equation}
Let us note a few additional peculiar features which come out of the
calculation \cite{long1}. The 4-point function%
\begin{equation}\label{X30}
G (u) := \lim_{|z_4| \to \infty} |z_4|^{4 \Delta_{10}} \left< \Phi_{10} (0)\epsilon (1) \epsilon (u) \Phi_{10}
(z_4) \right>
\end{equation}
 at $p=2$, i.e., for the
Ising model
is  
\begin{align}\label{X38b}
&G (u)\Big|_{p=2} = \frac{\Gamma(\frac{1}{3})^6}{27 \pi ^2} \frac{|u|^{\frac{2}{3}}}{|1-u|^{2}} \left| {}_2F_1\!\left({\textstyle -\frac{1}{3},\frac{2}{3};2;u}\right) \right|^{2}\nonumber \\
&\! +\! \frac{\Gamma(\frac{1}{3})^8}{54 \sqrt{3} \pi ^3}
\frac{|u|^{\frac{2}{3}}}{|1{-}u|^{2}} \Big[{}_2F_1({\textstyle
-\frac{1}{3},\frac{2}{3};2;u})G_{2,2}^{2,0}\Big(\overline{u}\left|
{\textstyle
\begin{array}{c}
 \frac{1}{3},\frac{4}{3} \\
 -1,0
\end{array}}
\right.\!\!\Big) \nn 
{+} c.c. \Big]
\end{align}
$G$ is the Meijer $G$-function, which has a logarithmic divergence at $u=0$,
\begin{equation*}\label{X37}
G_{2,2}^{2,0}\left(\overline{u}\left|
{\textstyle
\begin{array}{c}
 \frac{1}{3},\frac{4}{3} \\
 -1,0
\end{array}
}
\right.\!\!\right) = \frac{1}{3}\,\Gamma
({\textstyle \frac{1}{3}})^{2} \, {}_2F_1\!\left({\textstyle
-\frac{1}{3},\frac{2}{3};2;u}\right)  \ln (u) + \ldots\ ,
\end{equation*}
dropping regular terms. 
The structure of the result and the logarithmic divergence
remain valid for larger values of $p$, with the
parameters replaced by rational functions of $p$.
This behavior is consistent with 
the appearance of operators of canonical dimensions
$1$ and $0$ (different from the identity) in the OPE of 
$\epsilon$ with $\Phi_{1,0}$ as discussed in a similar case
in \cite{Gurarie93}. Logarithms are known to 
appear for operators 
on the boundary of the Kac table 
\cite{PearceRasmussenZuber} 
 and 
in disordered systems \cite{logdisorder}. 

Even more surprisingly, we attempted to perform the same
calculation for the fractal dimension of spin interfaces, using the
operator $\Phi_{01}$ as curve detector. 
There the
equivalent of (\ref{X30}) does not exist: At least we were not able to
construct a 4-point function, which satisfies the differential
equation induced by the 0-vector condition associated to $\epsilon
= \Phi_{12}$ at level 2, which is unique-valued, and reproduces the
correct OPE in the limit of $u\to 1$.



\begin{figure}
\Fig{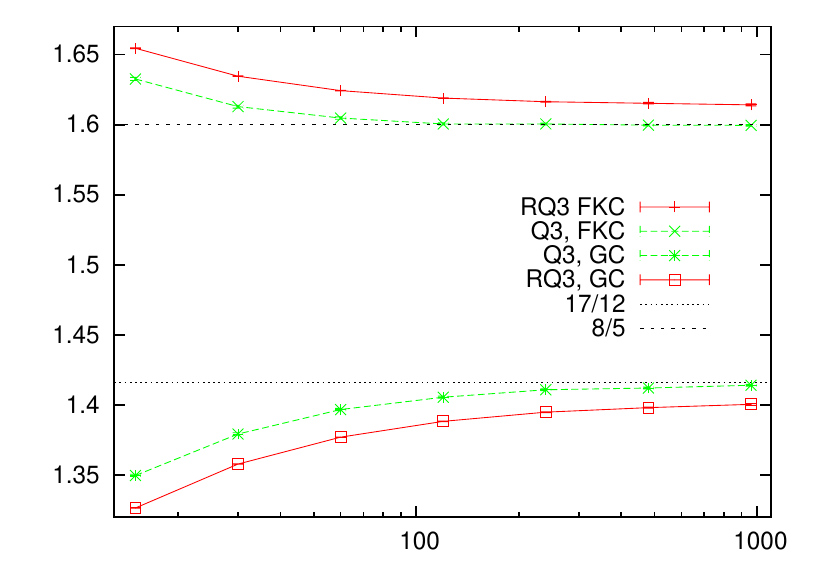}
\caption{Fractal dimension of FK and spin clusters, both for the pure
(green) and disordered (red) system, 
using the Wolff algorithm.}
\label{f:1}
\end{figure}

We now discuss our numerical results. For the Monte Carlo simulations
we use the Wolff cluster
algorithm \cite{Wolff1989}. It consists in randomly choosing a spin,
and then joining with probability ${\sf p}$ nearest-neighbor spins to belong
to the same cluster. This procedure is repeated until no nearest
neighbor can be joined anymore.  ${\sf p}$ itself is a quenched random
variable on the edges of the lattice, taken from the symmetric bimodal
distribution $\left\{{\sf p}_{1},{\sf p}_{2} \right\}= \left\{1-\exp (-\beta_{c}
  J_{1}), 1-\exp (-\beta_{c} J_{2}) \right\}$. The choice ${\sf p}_{1} {\sf p}_{2}
= q$ ensures that the system is at its critical point
\cite{KinzelDomany1981}. We use $J_{1}=J_{2}$ for the pure, and
$J_{1}/J_{2}=10$ for the random-bond disordered system. Once the
cluster is constructed, we assign to it with equal probability one of
the $q$ colors, as long as this is consistent with the boundary
conditions (BC) discussed below. Wolff has shown \cite{Wolff1989}
that this algorithm produces the correct statistical weight of the
Potts model, is ergodic, and that the critical slowing down is reduced
compared to ordinary Monte Carlo algorithms. In addition, it allows to
track both spin and FK clusters.  
In the simulations, we imposed boundary conditions which create a domain wall which span the lattice
as was already done by Gamsa and Cardy \cite{GamsaCardy2007}. We considered various types of 
conformally invariant boundary conditions: ``fluctuating'' ($a/\bar a$), ``fixed'' ($a/b$), and ``free'' ($a/\text{free}$), all of them 
giving the same result in the large-size limit. 

We measured the fractal dimension from the average length $l$ of the
domain wall as a function of the linear size $L$ of the lattice
$\overline {\langle l \rangle} \simeq L^{d_{\mathrm{f}}}$, where $\langle \cdots
\rangle$ denotes the thermal average and $\overline {\cdots}$
the disorder average \cite{footnotemoments}. The results presented here are
obtained with a thermal average over $\simeq 10^6 \times \tau$ for
the pure system and a disorder average over $\simeq 10^5$
configurations for the disordered system. $\tau$ is the
autocorrelation time which was first determined for each size, see
\cite{long1} for  details. Our simulations show that for the pure
system, all these domain walls have asymptotically the same fractal
dimension, with the exception of the {\em common} domain wall for
fixed BC, which has dimension one. 
In Fig.~\ref{f:1} we plot the
effective fractal dimension versus $L$. 
As $L\to\infty$ the fractal dimensions of the pure system converge to the
values predicted by conformal field theory,
$d_{\mathrm{f}}^{\mathrm{spin}} = \frac{17}{12}$, and
$d_{\mathrm{f}}^{\mathrm{FK}} =\frac{8}{5}$, corroborating partial
results by Gamsa and Cardy \cite{GamsaCardy2007}. Our estimate from
all BC, extrapolated to an infinite system gives
$d_{\mathrm{f}}^{\mathrm{spin}} = 1.416 \pm 0.002$, and
$d_{\mathrm{f}}^{\mathrm{FK}} = 1.599 \pm 0.002$.  For the disordered
system we find
\begin{equation}\label{}
d_{\mathrm{f}}^{\mathrm{spin}} = 1.401 \pm 0.003 \ ,
\qquad d_{\mathrm{f}}^{\mathrm{FK}} =1.614 \pm 0.003
\label{20}
\end{equation}
This is in excellent agreement with our analytical result
(\ref{16}). We have also checked the SLE duality relation $\kappa
\kappa' =16$. Using $d_{\mathrm{\kappa}} = 1+ \frac{\kappa}{8}$, we
find for the pure system $\kappa^{\mathrm{spin}} = 3.328 \pm 0.016$,
$\kappa^{\mathrm{FK}} = 4.792 \pm 0.016$, and $\kappa^{\mathrm{spin}}
\kappa ^{\mathrm{FK}}= 15.95 \pm 0.13$.  For the disordered system, we
find $\kappa^{\mathrm{spin}} = 3.208 \pm 0.024$, $\kappa^{\mathrm{FK}} =
4.912 \pm 0.024$, and $\kappa^{\mathrm{spin}} \kappa ^{\mathrm{FK}}= 15.76 \pm 0.20$.

\begin{figure}
\Fig{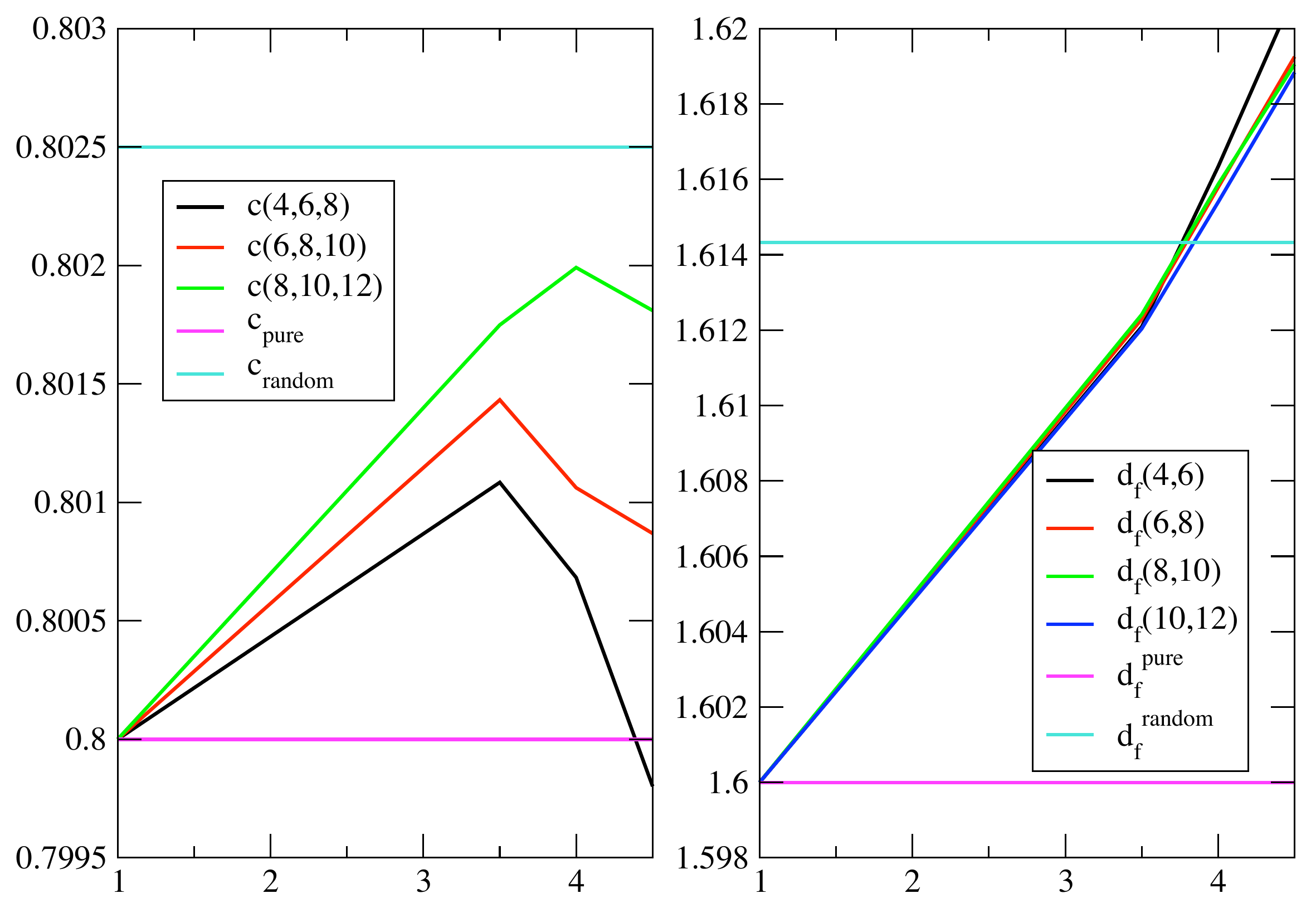}
\caption{Effective central charges $c(L)$ and fractal dimensions
  $d_{\rm f}^{\rm FK}(L)$ versus disorder strenght $s$, for
  $s=3.5,4.0,4.5$.  The linear interpolation is only a guide to the
  eye. Each curve has been normalized so that finite-size effects are
  absent for $s=1$ (no disorder). Horizontal lines give corresponding exact (resp.\
  perturbative) values for the pure (resp.\ disordered) system.}
\label{f:2}
\end{figure}

In the transfer-matrix approach, we studied the FK clusters
in the equivalent loop formulation \cite{DJLP99}. The loops are
defined on the medial lattice as the external and
internal  hulls of the FK clusters. 
The random
bonds were again drawn from a bimodal distribution, with an equal number of strong and weak bonds \cite{JP00}. The strength of the
disorder is conveniently characterized by the parameter $s$
\cite{JJmulti} defined by $J_1/J_2 = \ln(1+s
\sqrt{q})/\ln(1+\sqrt{q}/s)$, with $q=3$.
For a given fixed realization of the random bonds on long cylinders of
length $M=10^5$ and circumference $L=4,6,\ldots,12$ lattice spacings
(for the medial lattice) the corresponding free energy $f_j(L)$,
normalized per lattice site, was computed from the leading Lyapunov
exponent of the corresponding product of random transfer matrices.  The
transfer direction was taken axial with respect to the medial lattice
(hence diagonal with respect to the original square lattice supporting
the Potts spins) \cite{DJLP99}. Three different topological sectors
were considered, corresponding to enforcing $j=0,2,4$ loop segments to
propagate along the length direction of the cylinder.  The
fluctuations of these free energies were studied by averaging over at
least $M'=10^5$ independent cylinders. 

Conformal field theory predicts \cite{Cardy86} that $\overline{f_0(L)}
= \overline{f_0(\infty)} - \frac{\pi c}{6 L^2} + \frac{A}{L^4} +
\cdots$, where $c$ is the effective central charge and $A$ a
non-universal constant.  Applying this to three consecutive $L$ gives
estimates $c(L-4,L-2,L)$ shown in the left panel of Fig.~\ref{f:2}.
The fixed-point value $s^*$ of the disorder strength corresponds to
the locus of the maximum of $c$, and is estimated as $s^* = 4.0 \pm
0.3$ (using also data not shown here), improving on the value $s^* =
3.5 \pm 0.5$ reported earlier \cite{JJmulti}. The effective central
charge of the disordered model is estimated as $c(s^*) = 0.8024 \pm
0.0003$, in excellent agreement with the three-loop perturbative
result \cite{DotsenkoPiccoPujol1995} $c \simeq 0.8025$.

Correlation functions $G_j(M)$ are defined as the probability of
having $j$ loop segments propagate over a distance $M$ along the
cylinder axis without joining up. They are related to the free energy
gaps through $\Delta f_j(L) \equiv f_j(L)-f_0(L) = \frac{-1}{ML} \ln
G_j(M)$.  Their disorder-averaged $n$'th moment can be extracted from
the cumulant expansion \cite{randompotts} $\ln \overline{(G_j)^n} = n
\overline{\ln G_j} + \frac12 n^2 \overline{(\ln G_j -
  \overline{\ln G_j})^2} + \cdots$, where the quantities on the r.h.s.\ are self-averaging. Only the first two cumulants
contribute significantly.
CFT predicts \cite{Cardy84} that $\frac{-1}{ML} \ln \overline{(G_j)^n}
= \frac{2 \pi x_j}{L^2} + \frac{B}{L^4} + \cdots$, where the  $n$-dependent conformal
weights $x_j$ are related to the desired (multi)fractal dimensions via
$d_j = 2-x_j$. For  $n=1$, we have $d_2 = d_{\rm f}^{\rm FK}$
defined above; $d_4$ gives the dimension of
``red bonds'' (whose removal disconnects a cluster).
As seen from the right panel of Fig.~\ref{f:2}, the effective values of
$d_j$ depend strongly on $s$, so accuracy for $s^*$
is important \cite{JJmulti}. Using $s^* = 4.0 \pm 0.3$ we
estimate $d_2 = 1.615 \pm 0.002$, in excellent agreement with
(\ref{16}) and (\ref{20}).

To conclude, our analytical and numerical results for the fractal dimension of FK
domain walls agree well. Fractal dimensions of spin
interfaces  have been determined from numerics and
seem in agreement with the duality relation suggested by
SLE. Pending questions under investigation are possible multiscaling,
the fractal dimensions of spin interfaces and SLE type observables.

\acknowledgments We thank M.\ Bauer, D.\ Bernard, Vl.\ Dotsenko, A.\ Ludwig, T.\ Quella and P.\ Wiegmann for valuable discussions.  Supported by ANR (05-BLAN-0099-01).


\begin{thebibliography}{10}

\bibitem{schramm00}
O. Schramm, 
Israel J. Math. {\bf 118}, 221, (2000).

\bibitem{sleBernardBauerRev06}
Reviews: M. Bauer and D. Bernard, 
Phys. Rep.
{\bf 432} (2006) 115. 
J. Cardy, 
Ann. Phys. {\bf 318} (2005) 81. 

\bibitem{nonminimal}
R.~Santachiara, Nucl. Phys. B {\bf 793} (2008) 396; 
M.~Picco and R.~Santachiara,  Phys. Rev. Lett. {\bf 100} (2008)   015704.

\bibitem{turbulence}
D. Bernard et al. Nature Physics {\bf 2} 124 (2006).

\bibitem{spinglass}
D. Bernard, et al., 
Phys. Rev. B {\bf 76} (2007) 020403 (R) .


\bibitem{Ludwig1987}
A.W.W. Ludwig,
\newblock Nucl. Phys. B {\bf 285} (1987)   97;
A.W.W. Ludwig and J. Cardy, {\em ibid.}, 687.

\bibitem{DotsenkoPiccoPujol1995}
Vl.S.~Dotsenko, et al., 
\newblock Nucl. Phys. B {\bf 455} (1995)   701. 

\bibitem{randompotts}
M.~Picco, Phys.~Rev.~Lett.~{\bf 79} 2998 (1997);
J.~Cardy and J.L.~Jacobsen, Phys.~Rev.~Lett.~{\bf 79}, 4063 (1997);
J.L.~Jacobsen and J.~Cardy, Nucl.~Phys.~B {\bf 515}, 701 (1998).


\bibitem{Ludwig1990_DDP_97_Lewis98}
A.W.W. Ludwig, Nucl.~Phys.~B {\bf 330} (1990) 639; 
Vik.S.~Dotsenko, et al., 
Nucl.\ Phys. B {\bf 520} (1998) 633;
M.A.~Lewis, Europhys. Lett. {\bf 43} (1998) 189.

\bibitem{Wolff1989}
U.~Wolff,
\newblock Phys. Rev. Lett. {\bf 62} (1989)   361.

\bibitem{DotsenkoCFT}
Vl.S.~Dotsenko,
\newblock 
cel.archives-ouvertes.fr/cel-00092929



\bibitem{Nienhuis1984}
B.~Nienhuis,
\newblock J. Stat. Phys. {\bf 34} (1984)   731.

\bibitem{RushkinBettelheimGruzbergWiegmann2007}
I.~Rushkin, et al., 
\newblock J. Phys. A {\bf 40} (2007)   2165;
%
H.W.J. Bl\"ote, et al., 
\newblock Phys. Rev. Lett. {\bf 68} (1992)   3440.
%


\bibitem{KinzelDomany1981}
W.~Kinzel and E.~Domany,
\newblock Phys. Rev. B {\bf 23} (1981) 3421.

\bibitem{GamsaCardy2007}
A.~Gamsa and J.~Cardy,
\newblock J. Stat. Mech. (2007)   P08020.

\bibitem{long1}
Present authors, in preparation.

\bibitem{Gurarie93}
V.~Gurarie,
\newblock Nucl. Phys. B{\bf 410} (1993)   535.

\bibitem{PearceRasmussenZuber}
P.A.~Pearce, et al., 
\newblock J. Stat. Mech. {\bf 0611} (2006)   P017.


\bibitem{DJLP99}
Vl.S.~Dotsenko, et al., 
Nucl. Phys. B {\bf 546}, 505 (1999).

\bibitem{JJmulti}
J.L.~Jacobsen, Phys.~Rev.~E {\bf 61}, R6060 (2000).

\bibitem{JP00}
J.L.~Jacobsen and M.~Picco, Phys. Rev. E {\bf 61}, R13 (2000).

\bibitem{Cardy86}
H.W.J.~Bl\"ote, et al.\ 
Phys.~Rev.~Lett.~{\bf 56}, 742 (1986);
I.~Affleck, Phys.~Rev.~Lett.~{\bf 56}, 746 (1986).

\bibitem{Cardy84}
J.~Cardy, J.~Phys.~A {\bf 17}, L385 (1984).

\bibitem{logdisorder}
J. Cardy, 
cond-mat/9911024; S. Caux, 
hep-th/9511134.

\bibitem{footnotemoments} 
Measurements of the moments $\overline{ \langle  l \rangle ^m }$ for $m=1,2,\ldots$ are consistent with
an unambiguous definition of $d_{\mathrm{f}}$. 


\end{thebibliography}
\end{document}